\documentclass{ws-ijmpd}
\usepackage{bm}
\usepackage{latexsym}
\usepackage{dcolumn}
\usepackage{amsfonts,amssymb}
\usepackage{graphicx,epsfig}
\usepackage{psfrag}

\def\beq{\begin{equation}}
\def\eeq{\end{equation}}    
\def\eeq{\end{equation}}
\def\br{\begin{eqnarray}}
\def\er{\end{eqnarray}}
\def\ben{\begin{enumerate}}
\def\een{\end{enumerate}}
\def\bei{\begin{itemize}}
\def\eei{\end{itemize}}

\def\l{\left}
\def\r{\right}

\def\ast{{\em \/ Astron.\ J.~}}

\def\cqg{{\em \/Class.\ Quant.\ Grav.~}}

\def\grg{{\em \/Gen.\ Rela.\ Gravi.~}}

\def\mpla{{\em \/Mod.\ Phys.\ Lett.\ A~}}

\def\jhep{{\em \/ JHEP~}}

\def\plb{{\em  \/Phys.\ Lett.\ B~}}

\def\prd{{\em  \/Phys.\ Rev.\ D~}}
\def\prl{{\em  \/Phys.\ Rev.\ Lett.~}}

\begin{document}
\markboth{S. Shankaranarayanan and N.Dadhich}
{Non-singular black-holes on the brane}
%
\catchline{}{}{}{}{}
%

\title{\Large Non-singular black-holes on the brane}

\author{{S. Shankaranarayanan} \footnote{Present Address:
HEP Group, International Centre for Theoretical Physics,
Strada costiera 11, 34100 Trieste, Italy.} }
\address{DCTD, University of Azores, 9500 Ponta Delgada, Portugal.\\
Email: shanki@ictp.trieste.it} 

\author{ Naresh Dadhich}
\address{Inter University Center for Astronomy and Astrophysics,\\ 
Post Bag 4, Ganeshkhind, Pune 411 007, India.\\
Email: nkd@iucaa.ernet.in}

\maketitle

\begin{history}
\received{17 03 2004}
\revised{xx yy zz}
\end{history}

\begin{abstract}  
We investigate the possibility of obtaining non-singular black-hole
solutions in the brane world model by solving the effective field
equations for the induced metric on the brane. The Reissner-Nordstrom
solution on the brane was obtained by Dadhich et al by imposing the
null energy condition on the 3-brane for a bulk having non zero Weyl
curvature. In this work, we relax the condition of vanishing scalar
curvature $R$, however, retaining the null condition. We have shown
that it is possible to obtain class of static non-singular spherically
symmetric brane space-times admitting horizon.  We obtain one such
class of solution which is a regular version of the Reissner-Nordstrom
solution in the standard general relativity.
\end{abstract}



\section{Introduction}

General relativity provides an accurate description of a wide range of
classical gravitational phenomena. However, the description breaks
down at the space-time singularities.  According to the
Penrose-Hawking theorems \cite{Haw-Ellis} (see also
Ref. \refcite{senovilla-rev}), the manifolds arising as solutions of the
Einstein field equation in general relativity (GR) are, in general,
geodesically incomplete and thereby indicating the occurrence of
space-time singularities like the big-bang singularity in the
Universe's distant past or the singularity at the center of a
black-hole. The two key assumptions of the singularity theorems are:
(i) the matter must satisfy strong energy condition ($\rho\ge0, \rho +
\Sigma_i p_i\ge0$; SEC) and (ii) space-time must satisfy appropriate
causality conditions.

It is now a well accepted fact that a proper understanding of these
singularities or their avoidance may require new physics
\cite{horo-myers}. There have been various attempts in the literature,
in this direction, for example the scalar-tensor theories of gravity
in four (and other) dimensions \cite{scalar-tensor} and the higher
derivative gravity \cite{higher-deriv}. For the black hole
singularity, the focus is two fold: (i) the {\it internal structure of
black-holes}, the issues relating to mass-energy singularity and
non-unitary black-hole dynamics, and (ii) to understand the final
state of the gravitational collapse from some initial
configurations. There has been quite a vigorous activity on the latter
aspect for examining the cosmic censorship conjecture (see for a
recent review, see Ref. \refcite{pen1}); i.e. is the end product black
hole covering the singularity or the singularity naked?  Unfortunately
this study throws no light on the former aspect, the nature of the
singularity.

Besides these approaches there have been attempts to look for
non-singular black-hole solutions within the framework of general
relativity relaxing the energy conditions on the matter fields
\cite{GR-nonsing}. As mentioned earlier, the singularity theorems
require the stress tensor of the matter to satisfy SEC. The attempts
in this direction have been to look for black-hole solutions in which
the matter fields satisfy weak energy condition ($\rho\ge 0, \rho +
p_i\ge 0$; WEC). [Note that, from the energy-conservation theorem
\cite{Haw-Ellis}, it can be shown that the regular black-holes can not
exist for the matter satisfying dominant energy condition. Matter
fields satisfying weak energy condition implies null energy
condition.]

The idea of replacing the black-hole singularities by non-singular
vacuum cores traces back to Bardeen \refcite{bardeen}. In this model,
Bardeen assumed a 4-D spherically symmetric line-element of the form
\beq 
ds^2=\l(1-{2 R_g(r) \over r}\r) dt^2- 
\l(1-{2 R_g(r) \over r}\r)^{-1} dr^2 -r^2 d\Omega^2,
\label{eq:metric-Irina}
\eeq 
where 
\beq
R_g(r) = \frac{M \, r^3}{2 (r^2 + q^2)^{3/2}} 
\eeq
where $M$ is the mass of the black-hole and $q$ is the charge.  For
$q^2 < (16/27) M^2$, the line-element has an event horizon. It was
also shown that the space-time satisfies WEC. Various other authors
have also proposed regular black-hole solutions (for an incomplete list of
references, please see Ref. \refcite{GR-nonsing}).

Recently, Dymnikova \refcite{irina} has obtained a regular exact solution
(\ref{eq:metric-Irina}) to the Einstein equation, where
\br 
R_g(r)&=&\frac{r_g}{2}[1-\exp(-r^3/{r_1}^3)]; \nonumber \\
{r_1}&=&({r_0}^2 r_g)^{1/3};\quad {r_0}^2={3 \over \Lambda} = l^2;
\quad r_g= 2 M,
\label{eq:Irina-R}
\er
$\Lambda$ is related to the positive cosmological constant. The
space-time satisfies WEC. The above line-element has the following
interesting properties: (i) In the limit of $r \to 0$, the
line-element goes over to a de Sitter (ii) Asymptotically, the metric
goes over to a Schwarzschild  (iii) the metric has two
horizons -- event and cosmological.

The main aim of this study is to investigate non-singular black-hole
solutions in the large extra dimensional model, commonly known as the
brane world models. In the recent years, there has been a renewed
interest in models with extra dimensions in which the standard matter
fields are confined to the four-dimensional world (viewed as an
infinite hyper-surface -- brane) embedded in a higher dimensional
space-time (AdS bulk) where {\it only} gravity can propagate
\refcite{Randall-sundrum,rs,gt}. In these models, $5$-D bulk metric is
non-factorizable, and the small value of the true five-dimensional
Planck mass is related to its large effective four-dimensional value
by the large warp of the five-dimensional bulk. One of the features of
the brane models is that it reproduces Newtonian potential with higher
order corrections. These corrections contribute significantly in the
large curvature limits. (For a review on brane cosmology see, for
instance, Ref. \refcite{brane-cosmo}.)

Gravitational collapse on the brane has been studied by many authors
(for an incomplete list of references please see Refs.
\refcite{brane-collapse,naresh-roy}). In Ref. \refcite{naresh-roy}, the
authors obtain an exact black-hole solution of the effective Einstein
equation on the brane under the condition that the bulk has non zero
Weyl curvature and the brane space-time satisfies the null energy
condition. The solution is given by the usual Reissner-Nordstrom (RN)
metric where the charge parameter is thought of as a {\it tidal charge}
arising from the projection of the free gravity (the Weyl curvature)
of the bulk onto the brane. RN metric has thus been interpreted as
describing a black hole on the brane where electric charge's role is
taken over by the tidal charge and it can be thought of as the
analogue of the Schwarzschild solution on the brane. The tidal charge
like the RN electric charge would generate $1/r^2$ term in the
potential while the high energy modification to the Newtonian
potential cannot be any stronger than $1/r^3$ \cite{rs,gt}.  The cause
for this disagreement is the presence of tidal charge which is the
measure of the bulk Weyl curvature. The main drawback of the solution
is that we do not know the corresponding bulk solution. It is however
agreed that RN metric is a good approximation to a black hole on the
brane near the horizon \cite{shiro}.  This solution however has a
singularity at the center.

In this work, we would like to look for a non-singular form of the
brane black-hole solution obtained in Ref. \refcite{naresh-roy} by
solving the effective field equations for the induced metric on the
brane. As in many other works, we focus only on the brane
equation. That would mean looking for non-singular version of the
usual RN metric.  We shall therefore obtain non-singular RN black-hole
solution in a $D$-dimensional spherically symmetric space-time which
directly corresponds to $D$-brane non-singular black-hole in
$(D + 1)$-dimensional bulk.
 
The RN solution \cite{naresh-roy} on the brane was obtained by
imposing the null energy condition on the 3-brane and assuming that
the bulk has non zero Weyl curvature. This gave the RN metric as the
unique exact solution of the brane equation. For obtaining
non-singular solution, we have to relax one of the conditions and so
we drop the condition of vanishing scalar curvature $R$ and however
retain the null condition.  The latter will ensure occurrence of
horizon which is however required as we wish to find non-singular
brane black hole solution. It turns out that it is possible to obtain
a class of static non-singular spherically symmetric brane space-times
admitting horizon.

The rest of the paper is organized as follows: In section (2), we
obtain a non-singular RN solution in $D$-dimensional spherically
symmetric space-time and discuss its properties. In
section (3), we discuss the implications of the non-singular RN
solution in the context of the brane-world model and present our
conclusions.

\section{Non-singular RN black-hole in a $D-$ dimensional space-time}

The line-element of a $D$-dimensional spherically symmetric space-time
is given by
\beq
ds^2 = g(r) dt^2 - f(r) dr^2 - r^2 d\Omega^2 \, .
\label{eq:-gen-DD}
\eeq
where $d\Omega^2$ is the $(D - 2)$-dimensional angular line-element.
If we assume that the stress tensors satisfy the equation of state of
an anisotropic perfect fluid, we then have
\beq
T^t_t = T^r_r; \quad T^{\theta_1}_{\theta_1} = T^{\theta_2}_{\theta_2} 
= \cdots = T^{\theta_{D-2}}_{\theta_{D - 2}} \, ,
\eeq
which implies $f(r) = 1/g(r)$. Under this assumption, Einstein
equations of motion are 
\br
\label{eq:Grr}
8 \pi T^t_t = 8 \pi T^r_r &=& - (D - 2) \frac{g'(r)}{2 r} + \frac{
(D - 2) (D - 3)}{2} \l(\frac{1 - g(r)}{r^2}\r) \\
8 \pi T^{\theta_i}_{\theta_i} &=& - \frac{g''(r)}{2} - (D - 3) \frac{g'(r)}{r}
+ \frac{(D - 4) (D - 3)}{2} \l(\frac{1 - g(r)}{r^2}\r) 
\nonumber \\
\label{eq:G-phi}
&=& - \frac{\nabla^2_r g(r)}{2} + (D - 4)
\l[- \frac{g'(r)}{2 r} + \frac{(D - 3)}{2} \l(\frac{1 - g(r)}{r^2}\r)
\r]
\er
where the prime denotes differentiation with respect to $r$ and $i$ runs 
from $1$ to $(D - 2)$. Integrating Eq. (\ref{eq:Grr}), we get
\beq
g(r) = 1 - \frac{4 R_g(r)}{(D - 2) r^{D -3}} \, ,
\label{eq:DD-metric-coef}
\eeq
where
\beq
R_g(r) = 4 \pi \int_0^r \rho(x) x^{D - 2} dx \, .
\label{eq:DD-main}
\eeq
From the above expression, it is straightforward to see that different
choices of $\rho$ will lead to different forms of $R_g(r)$. As
discussed in Introduction, our aim is to look for a $D$-dimensional
non-singular charged black-hole solution satisfying Einstein's
equations. Let us consider the following mass and charge distributions
\br 
\rho_{\rm matter}(r)& = & c_1\exp[- c_2 r^{D - 1}] \nonumber \\
\rho_{\rm charge}(r)& = & \frac{c_3}{r^{2(D - 2)}} \l(1 - \exp[- c_4 
r^{2 (D - 2)}]\r) - {\mathcal Z} c_3 c_4 \exp[- c_4 r^{2 (D - 2)}]
\label{eq:DD-density}
\er
where $c_i$'s are constants which need to be determined and ${\mathcal
Z}$ is a real number. The above form of density distributions are
regular over all $r$ \footnote{In order to see this, we take two
extreme limits: a) As $r \to \infty$ the leading order terms of the
charge and matter densities go as (i) $\rho_{\rm matter} \to 0$ and
(ii) $\rho_{\rm charge} \to c_3/r^{2(D - 2)} $. b) As $r \to 0$, we
have (i) $\rho_{\rm matter} \to c_1$ and (ii) $\rho_{\rm charge} \to
(1 - {\mathcal Z}) c_3 c_4$.}.  Substituting the above density
distributions in Eq. (\ref{eq:DD-metric-coef}), we get
\br
\label{eq:DD1-RN}
g(r) &=& 1- \frac{16 \pi}{(D - 2)(D - 1)} \frac{1}{r^{D - 3}} 
\frac{c_1}{c_2} \l[1 - \exp(-c_2 r^{D -1})\r] \nonumber \\
&+& \frac{16 \pi c_3}{(D - 2)(D - 3)} \frac{1}{r^{2 (D - 3)}}
\l[1 - \exp(-c_4 r^{2 (D - 2)})\r] \\
&+& \frac{16 \pi c_3 c_4^{1/[2 (D - 2)]}}{(D - 2) r^{D - 3}} 
\l[\frac{{\mathcal Z}}{2 (D - 2)} - \frac{1}{(D - 3)}\r] 
\l(\Gamma\l[\frac{D - 1}{2(D -2)}, c_4 r^{2 (D - 2)}\r] - 
\Gamma\l[\frac{D - 1}{2(D -2)}\r]\r) \nonumber 
\er
where $\Gamma$ is the Gamma function.  The following points need to be
noted regarding the above expression:
\renewcommand{\theenumi}{\roman{enumi}}
\begin{enumerate}
\item The metric coefficients are regular over the whole range of $r$.

\item The choice of density in Eq. (\ref{eq:DD-density}) satisfies
WEC. 

\item In the limit of $r \to \infty$, the above line-element has the 
form of $D-$dimensional Reissner-Nordstrom.  

\item In the limit of $r \to 0$, the above line-element reduces to de
Sitter or Anti-de Sitter for different choices of $c_i$'s. In the
standard GR, the central core can only be de Sitter. However, in the
case of brane gravity, the central core can either be de Sitter or
Anti-de Sitter. We will discuss this aspect in the next section.

	In Ref. \refcite{kao}, the author obtained a charged black-hole
solution with a charged de Sitter core. It was assumed that the
stress tensor of a charged material to have a uniform charge-to-mass
ratio over all $r$. This gave the $r$-dependence of the charged
density to be the same as that of the $r$-dependence for the mass
density. However, the above stress tensor does not satisfy WEC as $r
\to 0$. In our case, we do not assume any such form for the charge
density and as mentioned earlier, the stress tensor satisfies WEC all
through.

\item The last term in the RHS of the above expression vanishes near
$r = 0$. In the limit of $r \to \infty$, the leading order term is
$1/r^{D - 3}$. Using this fact, and setting ${\mathcal Z} = 2 (D -
2)/(D - 3)$ in the above expression, we get
\br
g(r) &=& 1- \frac{16 \pi}{(D - 2)(D - 1)} \frac{1}{r^{D - 3}} 
\frac{c_1}{c_2} \l[1 - \exp(-c_2 r^{D -1})\r] \nonumber \\
&+& \frac{16 \pi c_3}{(D - 2)(D - 3)} \frac{1}{r^{2 (D - 3)}}
\l[1 - \exp(-c_4 r^{2 (D - 2)})\r] \, .
\label{eq:DD-RN-gen}
\er
In the rest of this section, we will consider this reduced form for
mathematical simplicity. However, the results derived can be extended
to the general case of Eq. (\ref{eq:DD1-RN}).
\end{enumerate}
The constants $(c_i)$ in the above line-element can be determined by
taking the two limits --- $r \to \infty$ and $r \to 0$ --- and
identifying them with the mass ($M$), charge ($Q$) and the
cosmological constant ($\Lambda$). We, thus, have
\br
\frac{8 \pi c_1}{3}- \frac{Q^2}{\alpha^4 } \mp \frac{1}{l^2} &=& 0 
\nonumber \\
c_3 = (D - 2) (D -3)\frac{Q^2}{16 \pi} \nonumber \\
\pm \alpha^{(D - 3)} \alpha^{(D - 1)} - r_g l^2 \alpha^{(D - 3)} + Q^2 l^2 &=& 0,
\er
where, the lower(upper) sign corresponds to Anti(de Sitter) core, $c_2
= \alpha^{-(D - 1)}$ and $c_4 = \alpha^{-2 (D - 2)}$. Even though, the
solution to the above algebraic equations are complicated, one can
in-principle solve it. In the case of $D = 4$, we have
\beq
\alpha = \pm \frac{A^{1/2}}{2} \pm \frac{A^{1/4}}{2} \l[
\frac{2 r_g l^2}{A} - 1 \r]^{1/2}
\eeq
where 
\br
A &=& \pm \frac{2^{7/3} l r_g^{-1/2} Q^2}{3^{1/3}B} + 
\frac{\sqrt{r_g l^2} B}{2^{1/3} 3^{2/3}} \\
B &=& \l[9 \sqrt{r_g l^2} + r_g^{-3/2}  
\sqrt{81 l^2 r_g^4 \mp 768 Q^6}\r]^{1/3}
\er
In the limit of $Q \to 0$, we have $\alpha^3 = l^2 r_g$. This
coincides with the result of Ref. \refcite{irina}. As noted earlier, in
the case of standard general relativity $Q^2$ is always non-negative
and hence the center has a de Sitter core.

\section{Discussions and Conclusions}

In Ref.  \refcite{naresh-roy}, the analogue of the Schwarzschild solution
was obtained by solving the modified vacuum equation on the brane. The
bulk space-time was taken to have non zero Weyl curvature. In order to
solve the equation completely, the authors imposed null energy
condition on the 3-brane. Using this condition, it was shown that RN
metric is the unique exact solution on the brane. The null condition
ensures occurrence of horizon which is what is required for a black
hole solution, however, this condition need to hold {\it only} at the
horizon and {\it not necessarily} everywhere else. This is the
situation for the Einstein-Yang-Mills black-hole solutions in 4D
gravity where the null energy condition holds only at the horizon
\cite{hairy-bh}. Numerical black-hole solutions have been obtained in
which the solution is (i) RN at the horizon and (ii) Schwarzschild
asymptotically. For regularity, space-time would not (as in the
Schwarzschild case) be empty and hence the scalar curvature would in
general be non zero even though the stress tensor contributed by the
bulk Weyl curvature would be trace free. It is then possible to obtain
a class of static non-singular spherically symmetric brane space-times
admitting horizon. For a non-singular black hole on the brane, it is
enough to find a regular version of the RN metric which is taken to
describe a black hole on the brane. In the preceding section, we have
obtained a regular version of the RN metric in $D$-dimensional
space-time;i.e. a regular black-hole solution on $(D-1)$-brane. It
would be interesting to find a generalization in which the null energy
condition holds only at the horizon but not globally. The solution
would however be numerical as is the case for the Einstein-Yang-Mills
black hole.

Even though, the brane black-hole solution is analogous to that of the
RN solution in standard general relativity, the horizon and the
singularity structure of the two black-holes are quite different. It
depends on the sign of $Q^2$. For the RN solution of the standard
general relativity, $Q^2>0$ because electric field energy is positive
and it leads to the familiar two horizons. However, for the brane
black-hole solution it has been argued that $Q^2$ must be negative as
it arises from the free gravitational field (which should have
negative energy [\refcite{nar}]) of the bulk space-time
[\refcite{naresh-roy}]. It is expected that the black-hole on the brane
should in the low energy limit imply corrections to the Schwarzschild
solution but its singularity and horizon structure should however
remain undisturbed. This can only happen if $Q^2<0$.

The black hole core could have been de Sitter or anti de Sitter
depending upon $Q^2$ being positive or negative. Since this ambiguity
has been resolved in favor of $Q^2<0$, the black hole center would be
AdS. Asymptotically it approaches RN/Schwarzschild space-time while in
the interior it would go to AdS at the center. The line-element
(\ref{eq:-gen-DD}) with $g(r)$ given by Eq. (\ref{eq:DD-RN-gen})
describes a regular black-hole space-time free of singularity and it
admits like the Schwarzschild black-hole only one horizon.  Since
the requirement of regularity is not uniquely specifiable, the regular
solution so obtained would not however be unique. The field is
therefore quite open and it is only the physical properties that would
distinguish one solution from the other.

\section*{Acknowledgments}
\noindent
S. S would wish to thank Nuno Barros e Sa for useful discussions. S. S
gratefully acknowledges support from Funda\c c\~ao para a Ci\^encia e
a Tecnologia (Portugal) under the grant SFRH/BI/9622/2002.

\end{document}